# Parameters Affecting Temporal Resolution of Time Resolved Integrative Optical Neutron Detector (TRION)

**Mor I.[a1], Vartsky D.[a], Dangendorf V.[b], Bar D.[a], Feldman G.[a], , Goldberg M. B.[a], Tittelmeier K.[b], Bromberger *B.* [b], Brandis M. [a], Weierganz M. [b],**

[a] *Soreq NRC,*
*Yavne 81800, Israel*
[b] *Physikalisch-Technische Bundesanstalt (PTB),*
*38116 Braunschweig, Germany*
*E-mail*: ilanmor@yahoo.com

ABSTRACT: The Time-Resolved Integrative Optical Neutron (TRION) detector was developed for Fast Neutron Resonance Radiography (FNRR), a fast-neutron transmission imaging method that exploits characteristic energy-variations of the total scattering cross-section in the $E_n$ = 1-10 MeV range to detect specific elements within a radiographed object. As opposed to classical event-counting time of flight (ECTOF), it integrates the detector signal during a well-defined neutron Time of Flight window corresponding to a pre-selected energy bin, e.g., the energy-interval spanning a cross-section resonance of an element such as C, O and N. The integrative characteristic of the detector permits loss-free operation at very intense, pulsed neutron fluxes, at a cost however, of recorded temporal resolution degradation.

This work presents a theoretical and experimental evaluation of detector related parameters which affect temporal resolution of the TRION system.

[1] Corresponding author

## Contents

## 1. Introduction

The Time-Resolved Integrative Optical Neutron (TRION) detector was developed for Fast Neutron Resonance Radiography (FNRR) [1], a fast-neutron transmission imaging method that exploits characteristic energy-variations of the total scattering cross-section in the $E_n$ = 1-10 MeV range to detect specific elements within a radiographed object. FNRR holds promise for detecting a broad range of conventional and improvised explosives, due to its ability to determine simultaneously the identity and density distribution of the principal elements present in explosives, such as C, O and N.

The variant of FNRR with a pulsed neutron source, known as Pulsed Fast Neutron Transmission Spectroscopy (PFNTS), was proposed and first studied by the Oregon University group [2-4] for detection of explosives. The method was subsequently refined and taken through several blind tests for the FAA by Tensor-Technology, Inc. [5-7]. In the PFNTS method, a ns-pulsed, broad-energy (1-10 MeV) neutron beam is incident on the inspected object and the transmitted neutron spectrum is measured by the Time-of-Flight (TOF) technique. Both the Oregon University and Tensor groups employed the conventional Event-Counting TOF (ECTOF) spectroscopy mode, in which the elapsed time for an individual neutron to arrive at the detector following its creation in the target during the beam burst is recorded.

Both groups employed large-area detector arrays consisting of individual plastic scintillators (dimensions: several cm), each coupled to a photomultiplier tube via a light guide. The pixel size determined by these detectors posed an intrinsic limitation on the position resolution, which did not permit reliable detection of small and thin objects, such as thin-sheet explosives. Reduction of pixel size while using the above approach would have entailed an increase in the quantity of electronics at a prohibitive cost.

In order to respond to the above challenge, a new type of PFNTS detectors was developed in recent years – a Time Resolved Integrative Optical Neutron (TRION) detector. These detectors are capable of providing mm-size spatial resolution and good TOF spectroscopy (several ns) per pixel as well as the ability to operate at high neutron fluxes ($\geq 10^6$ n/(s×cm$^2$)).

As previously mentioned, TRION employs an integrative optical TOF (**ITOF**) technique. As opposed to event-counting time of flight (**ECTOF),** it integrates a neutron image during a well-defined neutron-Time-of-Flight window ($\Delta T_{TOF}$) corresponding to a specific energy bin, e.g., the energy-interval spanning a cross-section resonance. A $\Delta T_{TOF}$ is selected by the finite width of a gate signal which is delayed relative to the start of a neutron burst. The full TOF spectrum is obtained by varying the delay of the gate signal relative to the start time of the neutron burst [8-10]. In an integrative detector, information such as the energy deposited by each detected event and its exact arrival time is lost and only the integral information is recorded. Compared to existing ECTOF techniques, the advantages of this approach are: 1) it can provide excellent spatial resolution at an affordable cost and 2) it permits operation at very high neutron fluxes. The disadvantages are: 1) TOF resolution is dictated by gate width and detector response; 2) Sequential scanning of multiple TOF windows over many neutron pulses is required for accumulation of entire time spectrum and 3) a reduced signal-to-noise ratio [11].

The parameters which affect spatial-resolution with TRION have been reviewed in ref. [12]. The following presents a theoretical and experimental evaluation of detector-related parameters which affect the temporal resolution of the TRION system.

### 1.1 TRION concept

The TRION concept was first proposed in 2004 by the PTB group [13] and several prototypes were jointly developed and built at Soreq NRC and PTB [9-13]. Here we shall give only a brief outline of the two generations of the TRION detector that were developed. A detailed description of the TRION detectors can be found in [9].

The TRION detector is designed to detect fast-neutron pulses produced, for example, in the $^{9}$Be(d,n) reaction using a pulsed (~1-2 ns pulse width, 1-2 MHz repetition rate) deuteron beam. After a specific Time-of-Flight (TOF) that depends on the neutron energy and the distance between source and detector, the fast-neutrons impinge on the plastic scintillating fiber screen causing the emission of light from the screen surface, mainly via recoil protons. The light is reflected by a high reflectivity mirror, positioned at an angle of 45° relative to the neutron beam direction, towards a large aperture collecting lens positioned and subsequently focused on an image-intensifier (I-I). In the first generation of TRION, the latter not only amplifies the light intensity but, more importantly, acts as an electronic shutter that is opened for a gate period of $\Delta t$ at a fixed, pre-selected delay relative to each beam pulse. The entire TOF spectrum is obtained by shifting the delay of the gate relatively to the time of the beam burst. The exposure time control is performed by applying a fast (ns rise time) gate pulse to the photocathode of the I-I. This gate signal is a fast high voltage pulse (ca. -200 V amplitude) triggered by a computer controlled gate and delay generator (GD&G) unit. Repetition rate for the beam pulses and thus also for the gate pulses are up to 2 MHz. The screen of the I-I is viewed by a cooled CCD camera and images are integrated over many beam bursts, with acquisition times ranging from tens to several hundreds of seconds. All system components are mounted in a light-tight enclosure.

In TRION Gen. 1, the entire TOF spectrum is obtained by scanning the delay of the integrating gate relative to the start time of the neutron burst through the whole relevant neutron TOF spectrum.. This scan can be a lengthy process and in order to shorten it TRION Generation 2 (Gen. 2) detector was developed.

In TRION Gen. 2, seen in Fig. 1, the image transferred by the collecting lens is first amplified by an un-gated custom-made image-intensifier with a fast phosphor screen (decay time of the order of 2 ns [9]). This fast phosphor screen is viewed by four ns-gated CCD cameras. The fast gating element in this case is an 18 mm in diameter image-intensifier positioned in front of each CCD camera and gated seperately by a fast High-Voltage (HV) pulser. Each camera acquires a transmission image corresponding to a different energy region. Thus TRION Gen. 2 can scan the entire TOF (energy) region 4 times faster than TRION Gen. 1.

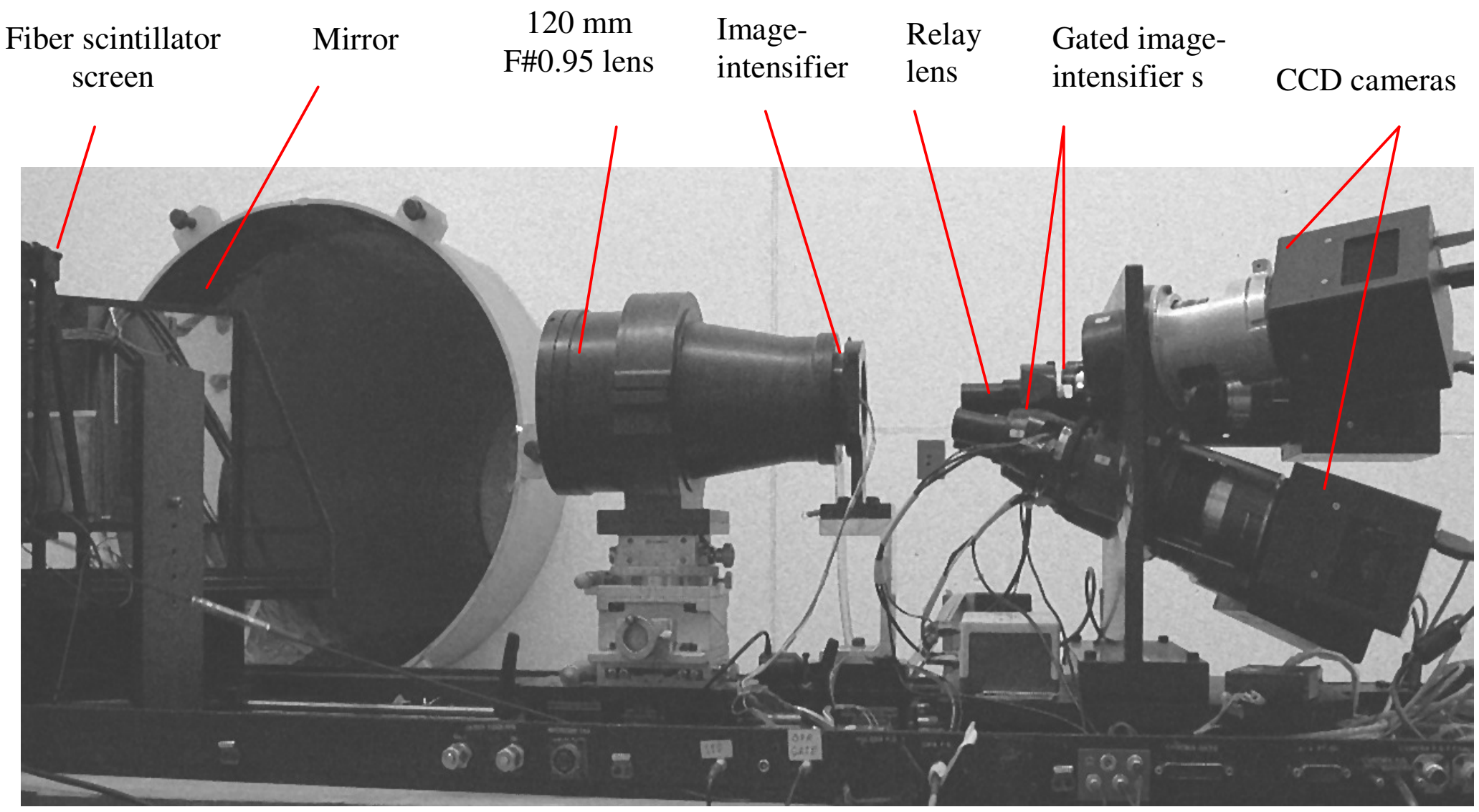

**Fig. 1 TRION Gen. 2**

Although the temporal resolution of TRION is adequate for most of our purposes, it is inferior to that obtainable with the classical event-counting TOF method.

Fig. 2 shows the transmission through 10 cm graphite block vs. time-of-flight (neutron energy) obtained with TRION (purple) and a non-imaging plastic scintillator detector operating in event counting mode (blue).

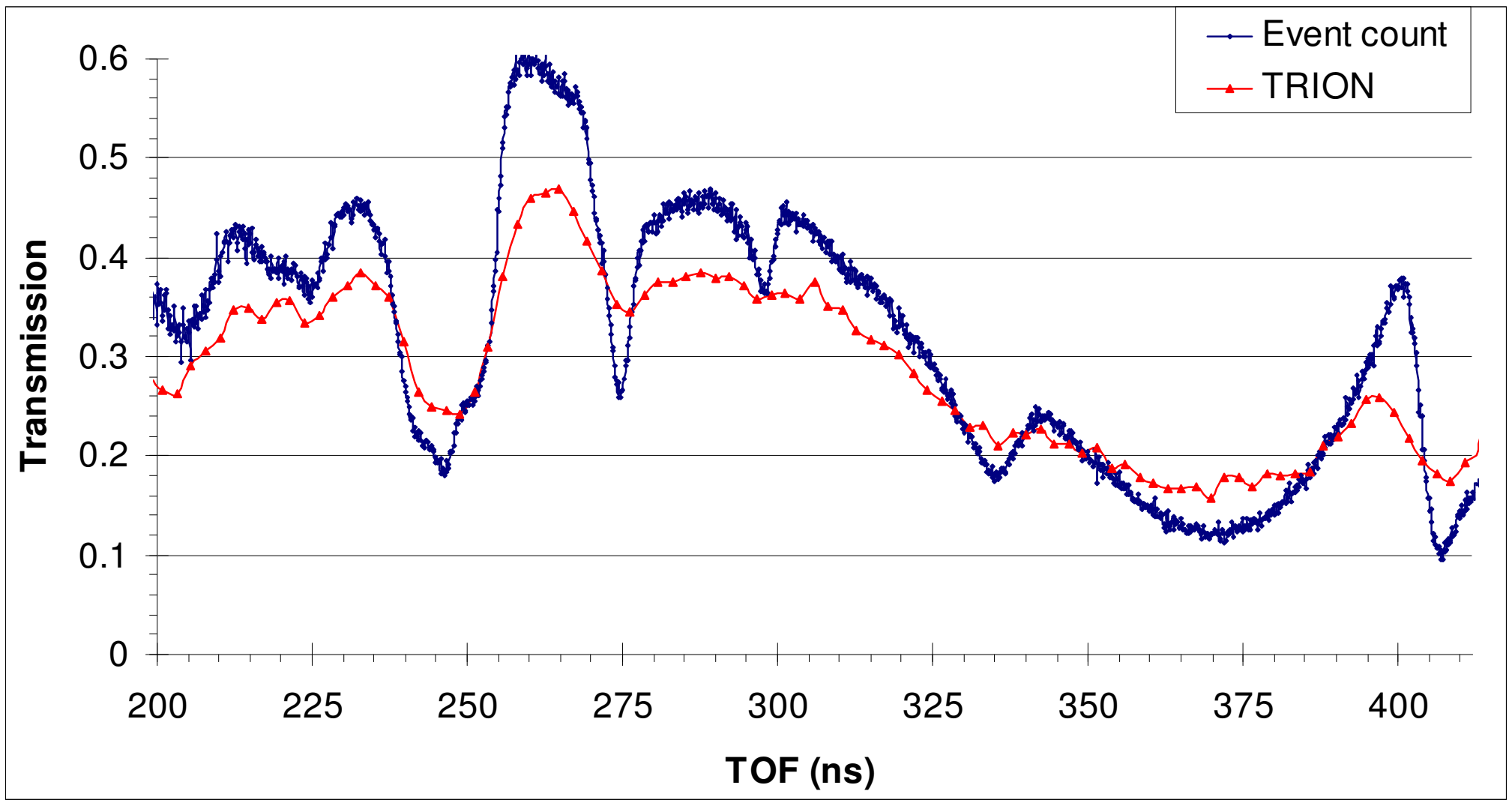

**Fig. 2 TOF spectra transmitted through 10 cm thick graphite block. Event count (blue), TRION (pink). Corresponding neutron energy range: 3 – 13 MeV**

As can be observed TRION suffers from considerable loss of temporal resolution, especially for sharp transitions.

The following chapter will examine the parameters that influence TRION's temporal resolution performance.

## 2. Parameters affecting temporal resolution in TRION detector

As opposed to the conventional event-counting variant of the TOF method, in which the arrival time of each detected neutron is recorded, TRION captures an image integrated over many neutron bursts, at a fixed time $\boldsymbol{t_{TOF}}$ with a width of $\boldsymbol{\Delta t}$ relative to the start time of each beam burst, which corresponds to a selected energy window around $\boldsymbol{E_n}$ determined by $\boldsymbol{\Delta t}$. In case of TRION Gen 2 there are four $\Delta t$ windows per beam burst as seen in Fig. 3.

Thus, in order to register scintillation light related exclusively to neutrons of a specific energy, TRION's Image Intensifier (I-I) needs to be time-gated. The temporal resolution, defined by the width of the gate pulse and the response speed of the I-I, directly affects the the sharpness of the features (peaks and dips) in the detected neutron TOF spectrum.

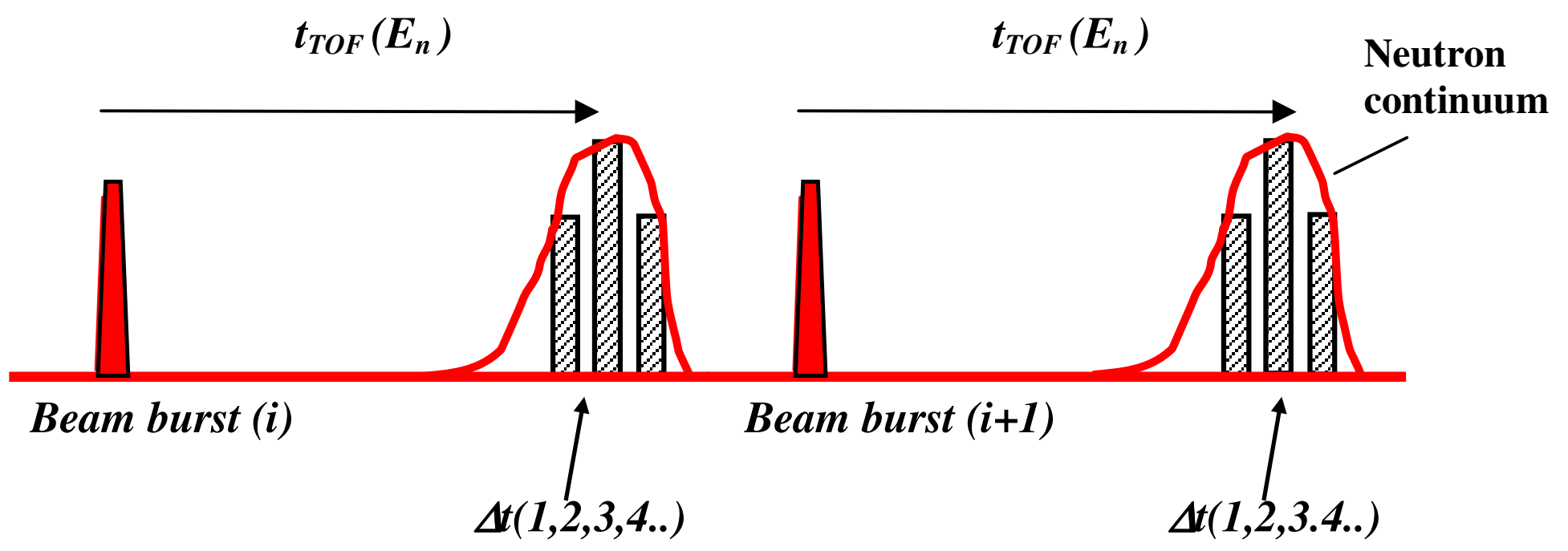

**Fig. 3 TRION image acquisition regime per neutron beam pulse: the image is acquired after time $t_{TOF}$ corresponding to neutron energy $E_n$**

The temporal resolution of TRION is governed by the following factors:

- Accelerator beam burst duration
- Neutron source-to-detector distance
- Gate and delay time jitter
- Thickness of the detector
- Scintillation decay time constants
- Life-time of scattered neutron inside the detector
- The minimal achievable I-I gate width

Due to the relatively large source-detector distances (D) employed (ca. 12 m), the detector thickness (*d=5* cm at maximum) plays a rather insignificant role in the temporal resolution of TRION ($\Delta t/t_{TOF} = d/D < 0.5\%$). To avoid time-jitter between the Gate & Delay Generator (G&DG) clock and the accelerator beam pulse, the gate & delay unit clock was phase-locked to the cyclotron frequency, reducing the time jitter to values below 200 ps.

In the following we shall try to determine separately the influence of the latter three factors (scintillation light decay time, life-time of scattered neutron inside the detector and I-I gate width). The two other factors which affect the temporal resolution, namely, accelerator beam burst duration and target-to-detector distance, are not detector-related and will not be addressed in this footing.

### 2.1 Effect of scintillator decay time

The temporal resolution and therfore resonant contrast is affected by the decay time constant of the scintillation light in TOF measurements perforemd by TRION.

Long scintillation decay time-constants can generate memory-effect from light created by faster neutrons, which reach the detector earlier than the neutrons of interest.Its effect on the time spectrum is expressed as a convolution of the real neutron arrival TOF spectrum with the scintillator light decay curve.

In order to determine the pure effect of the scintillator light decay curve we convolved a TOF spectrum obtained using a conventional event-counting method with a known light decay curve.

For this purpose we collected a TOF spectrum of neutrons transmitted through 10 cm graphite block using a thin NE102 plastic scintillator detector, coupled to a photomultiplier (2" in diameter), operating in an event counting mode. Accelerator pulse duration time was ca. 1.5 ns.

The decay curve of a NE102 plastic scintillator (composition similar to polystyrene) is taken from refs. [16-17] and shown in Fig. 4. It can be decomposed into 3 components with decay constants of approximately 2.5, 12 and 68 ns. The ratio of amplitudes of each component is 1: 0.026: 0.0023 respectively. This is considered to be a fast scintillator and has been chosen as the material used for our screen.

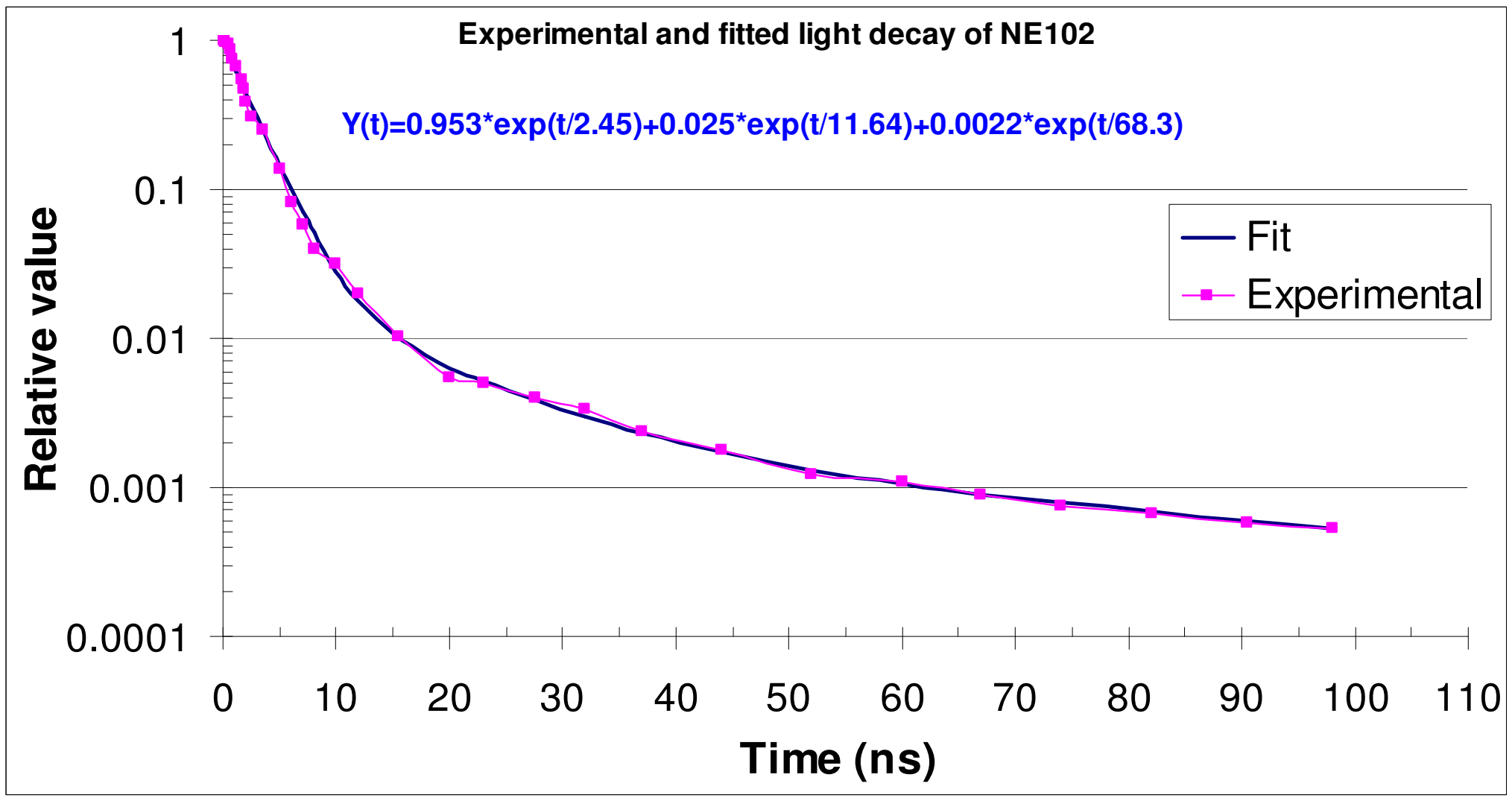

**Fig. 4 Experimental and fitted light decay of NE102 [16-17]**

Fig. 5 shows event-counting TOF spectrum (pulse duration time was ca. 1.5 ns) following transmission through 10 cm thick graphite as obtained by the NE102 scintillator (blue line), and the same spectrum after convolution (pink line) with the NE102 light decay curve of Fig. 4. As can be observed, the finite decay of the light reduces the contrast, especially for narrow resonances. The contrast was defined as the difference between the peak and the dip, normalized to the sum of these values. The reduction in contrast due to the finite decay time of the scintillator was **11%**, **57%** and **41%** for the time (energy) intervals indicated by **a**, **b** and **c** respectively.

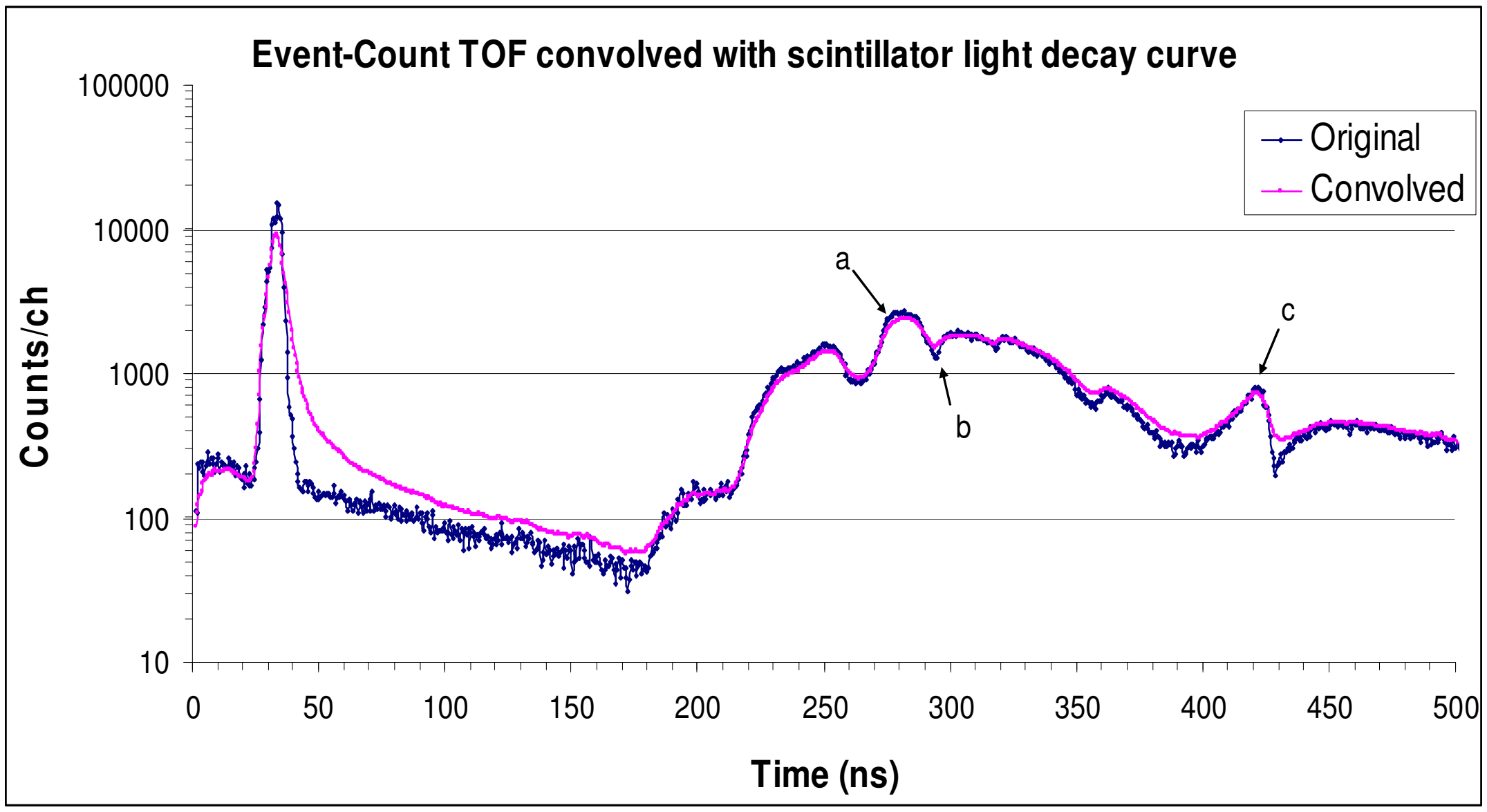

**Fig. 5 Event counting TOF (blue) and its convolution with the decay curve of NE102 (pink) for a 10 cm thick graphite absorber**

### 2.2 Minimal achievable image-intensifier (I-I) gate width

As stated previously, temporal resolution is principally determined by the image intensifier (I-I) gate-width. The shorter the gating time, the better the TOF resolution. As the diameter of the I-I photocathode increases, its capacitance and inductance increase and it becomes increasingly more difficult to gate for short time-intervals. The TRION Gen. 1 detector used a 40 mm diameter image intensifier and a custom-designed HV pulser. The pulser output switches the I-I photocathode voltage between +50V (intensifier-OFF) and -150V (intensifier ON). The first generation of TRION (TRION Gen. 1) achieved a minimal gating window of 20 ns for the TOF measurements. This is a long gating time which caused appreciable loss of temporal contrast, however, most of our work with Gen. 1 was done with this gating time. At later stages, the TRION Gen. 1 I-I gating window was shortened to about 11 ns. The effective gating window of the I-I was determined by scanning a ps-laser pulse through the gate window of the I-I tubes and measuring the relative intensity of the detected-amplified light (emitted from the intensifier phosphor) relative to the time difference between ps-laser pulse and rising edge of the gate pulse. (Further details on the determination process of the effective gating window can be found in refs. [9-10]).

TRION Gen. 2 used smaller image intensifiers and improved HV pulsers based on self terminated clipping line pulsers [9]. The shortest possible effective opening time of TRION Gen. 2 was 4.65 ns.

### 2.2.1 The effect of I-I gate width

In order to evaluate the influence of the I-I gate width on contrast, the event-counting spectrum described in previous sections was convolved with a square gate opening function using a 10 ns and 20 ns time window (corresponding to the gate-width at which the TRION detector was operated). The result of this convolution, seen in Fig. 6, shows that for the 10 ns window (green) the contrast reduction is expected to be **3%**, **32%** and **13%** for the a,b and c intervals indicated in Fig. 6. The 20 ns gate (red) caused a larger contrast reduction: **15%**, **37%** and **32%** respectively.

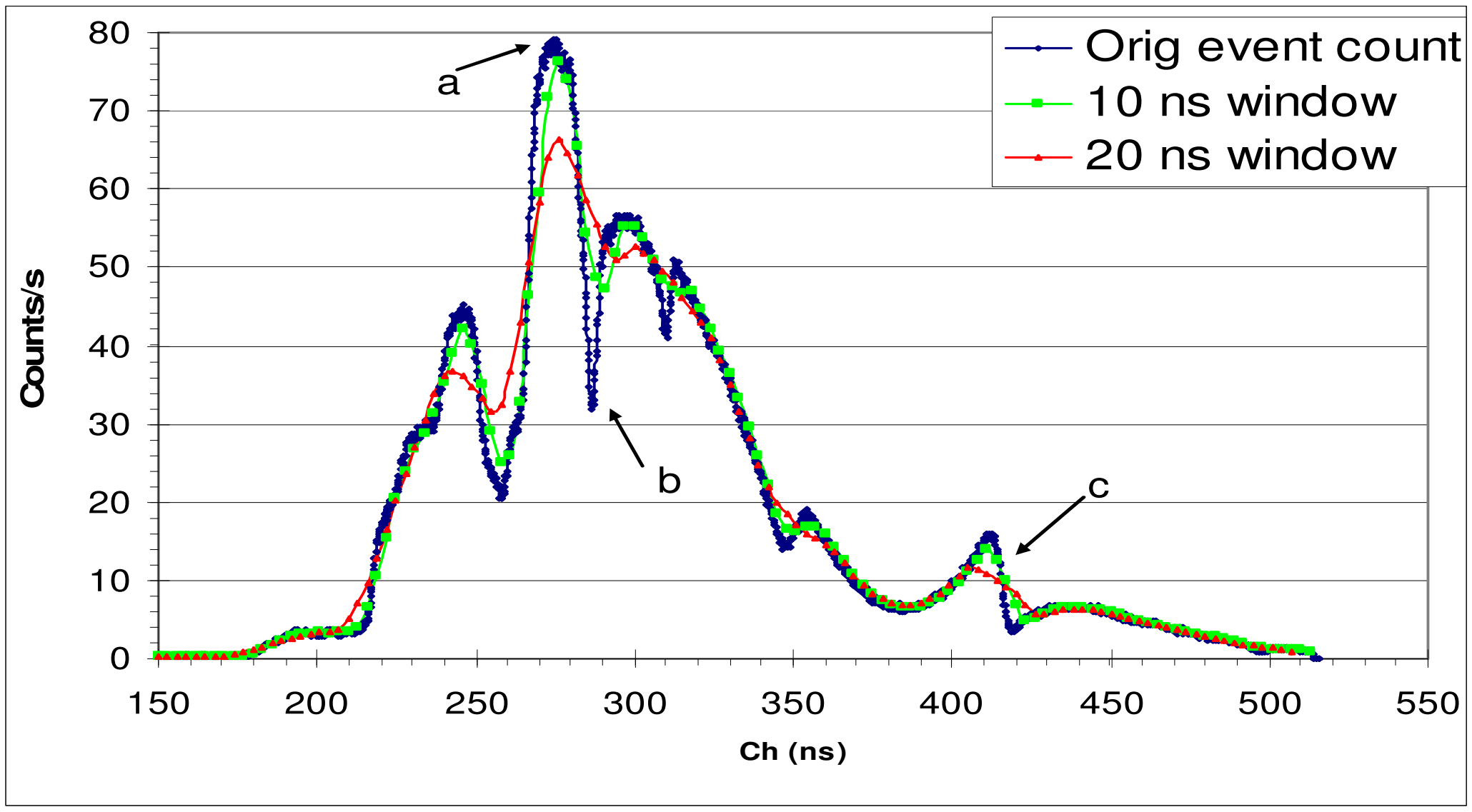

**Fig. 6 Effect of I-I gate width on TOF spectrum. Measured event-counting spectrum averaged using a 10 ns and 20 ns time window**

It would thus appear that scintillation decay time has a greater influence on contrast than the 10 ns gate width does; however, at longer gating periods such as 20 ns, it is the gate width that causes the most serious loss in contrast.

**2.3 Life-time of the scattered neutron within the scintillator**

An additional factor which may affect the temporal behavior of the detector is the survival time of neutrons scattered within the scintillating screen. If the screen has large dimensions, the probability for multiple scattering within it is significant. In such a case, the neutron may deposit its energy in the screen over a time interval comparable to longer than the characteristic times of the above mentioned processes.

The effect of the temporal behaviour of the residual light due to neutron scattering within the scintillaor screen was investigated by Monte-Carlo calculation for the 3 cm thick screen. The entire 20×20×3 $cm^3$ scintillating screen was irradiated with a broad beam of neutrons and the time behavior of the light is observed in the central area of the screen. Fig. 7 shows the amount of light per incident neutron vs. time (since the neutron had entered the scintillator screen) in the central pixel of the 3 cm thick screen for 2, 7.5 and 14 MeV neutrons.

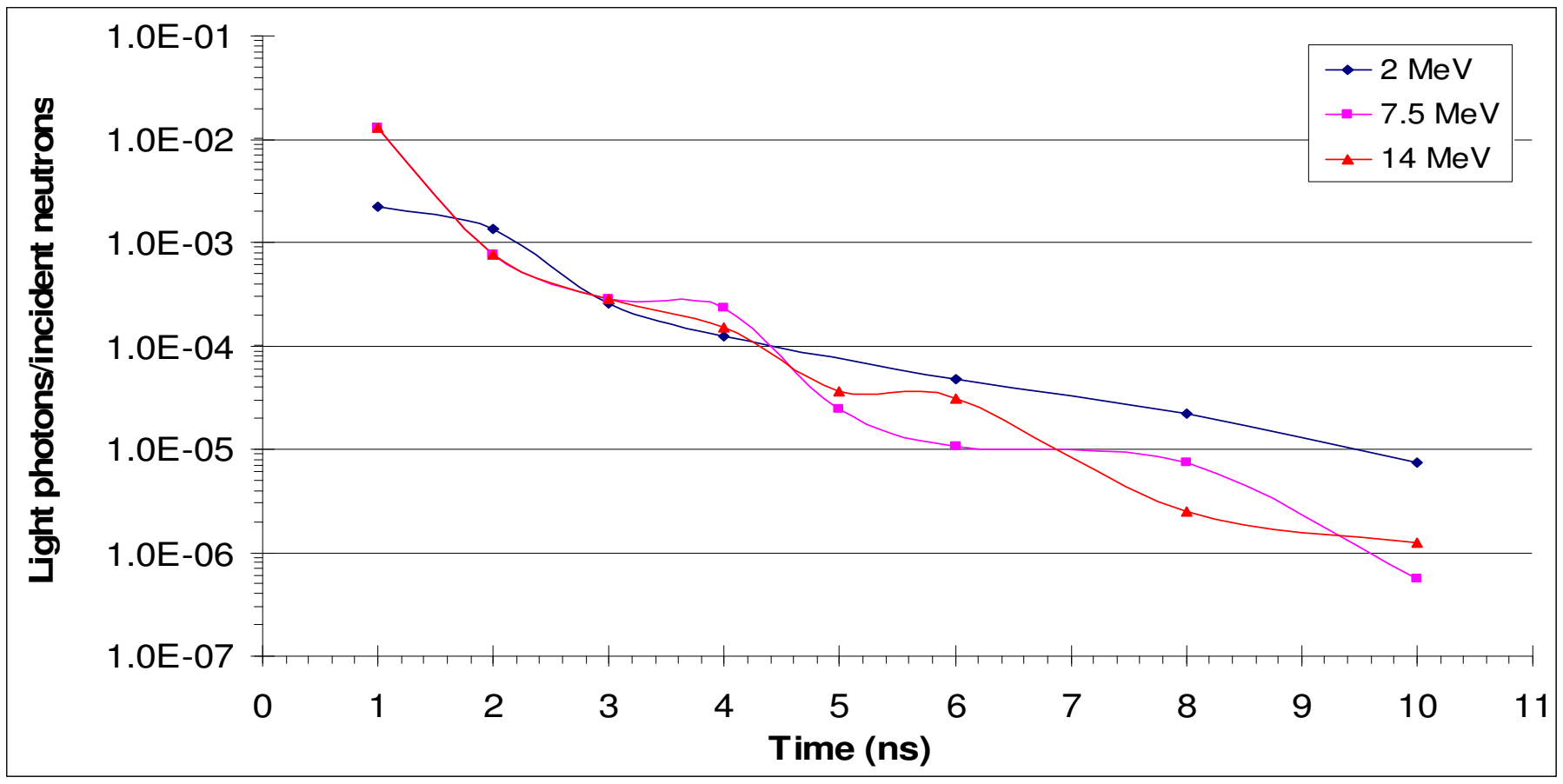

**Fig. 7 Light decay in the central pixel following neutron irradiation of the entire screen. (3 cm thick screen)**

For 2 MeV neutrons the light decays to 1/10 and 1/100 of its' value after 3 ns and 8 ns respectively, from the time of neutron arrival. For both higher neutron energies the corresponding times are approximately 1.8 ns and 4 ns.

As the time required for light to decay to 1/100 of its' value is within our gating time (4-6 ns), our conclusion is that for the 3 cm thick screen this effect is not of major importance; however, it should be taken into account and corrected for in larger screens.

Neutron scattering within the scintillator also affects the spatial resolution, as explained in ref. [12].

In relation to the above, it is interesting to calculate the spatial point spread function (PSF) as a function of time for several neutron energies. In this Monte-Carlo simulation, the central pixel of the 20×20×3 $cm^3$ scintillating screen was irradiated with a pencil-beam of neutrons. The time-and-space-history of each neutron were followed since the time it impinged on the screen.

The amount of scintillation light in each pixel due to this beam was determined and displayed, thus creating a scintillation image. A series of images were created vs. time. Fig. 8 shows the light images vs. time for a 1 MeV neutron beam (as 1 MeV neutrons will spend longer times within the screen than neutrons of higher energies).

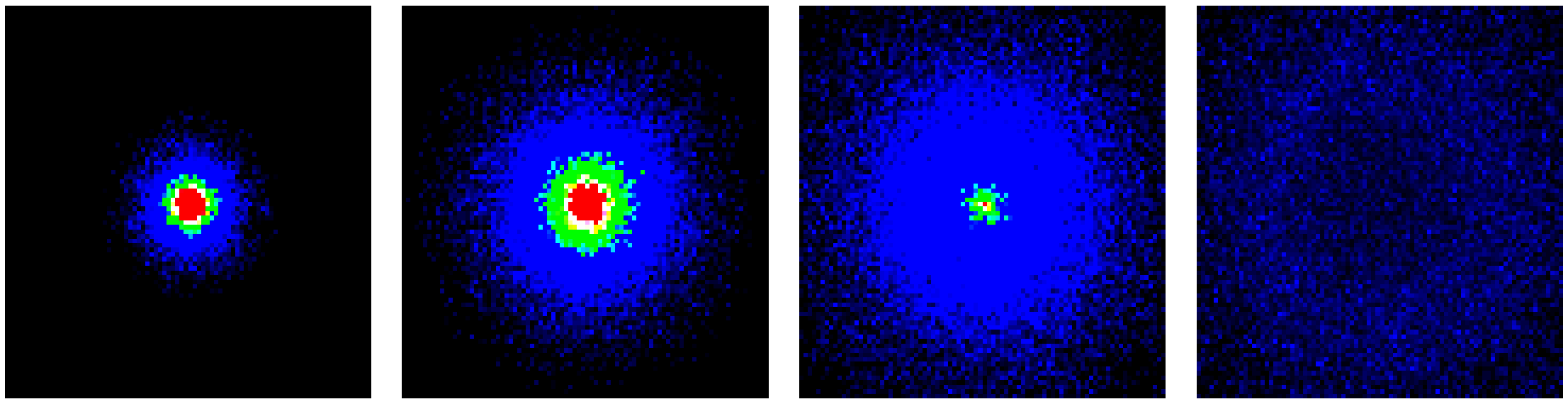

**1 ns** **2 ns** **3 ns** **4 ns**

**Fig. 8 Images of light created by 1 MeV neutrons vs. time since their arrival to the central pixel of the screen. Hottest color (red) indicates highest number of light photons**

Figures 9 a, b show the profiles of light taken through the central row of fibers in the screen for 1 and 7.5 MeV neutrons respectively.

The PSF broadens with time, after several ns - the time necessary for the neutrons to traverse the thickness of the screen, and it becomes a faint circular expanding wave. The time to reach this state decreases with neutron energy, as one would intuitively expect.

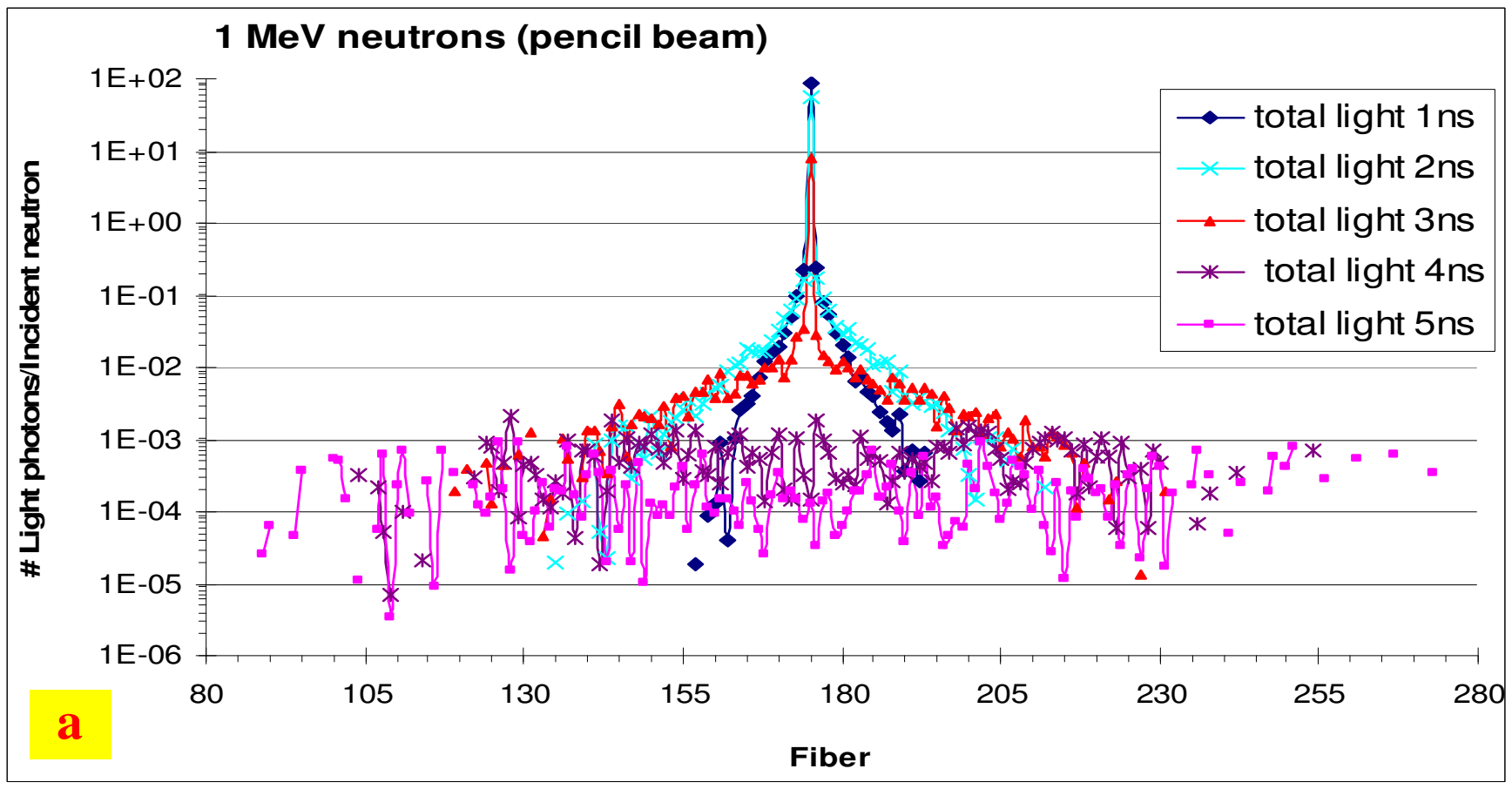

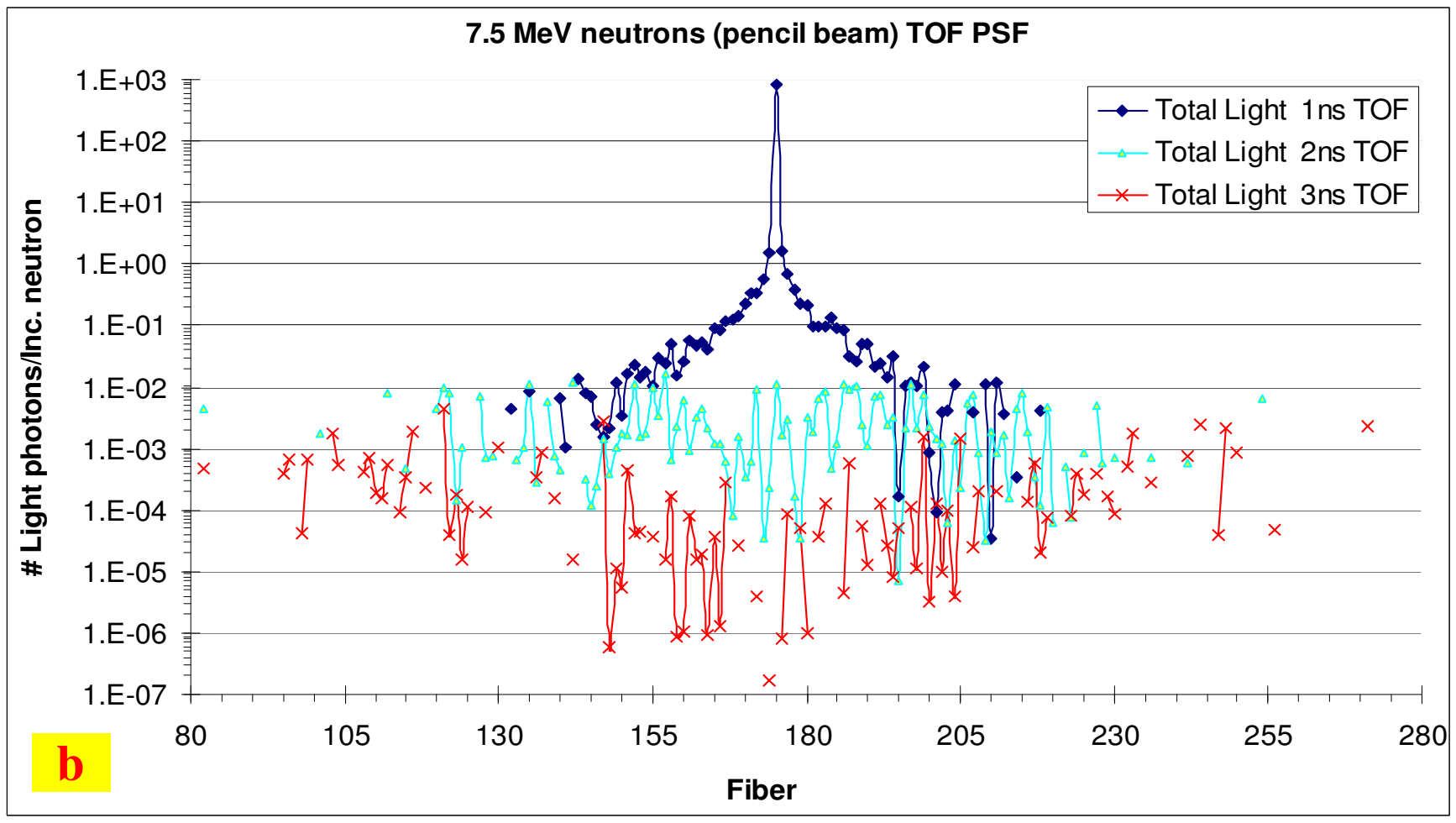

**Fig. 9 a,b) PSF vs. time for 1 and 7.5 MeV neutrons, respectively, for 3 cm thick screen**

For comparison, Fig. 10 shows a 7.5 MeV PSF for a 10 cm thick screen. As expected the neutrons spend more time in the larger screen.

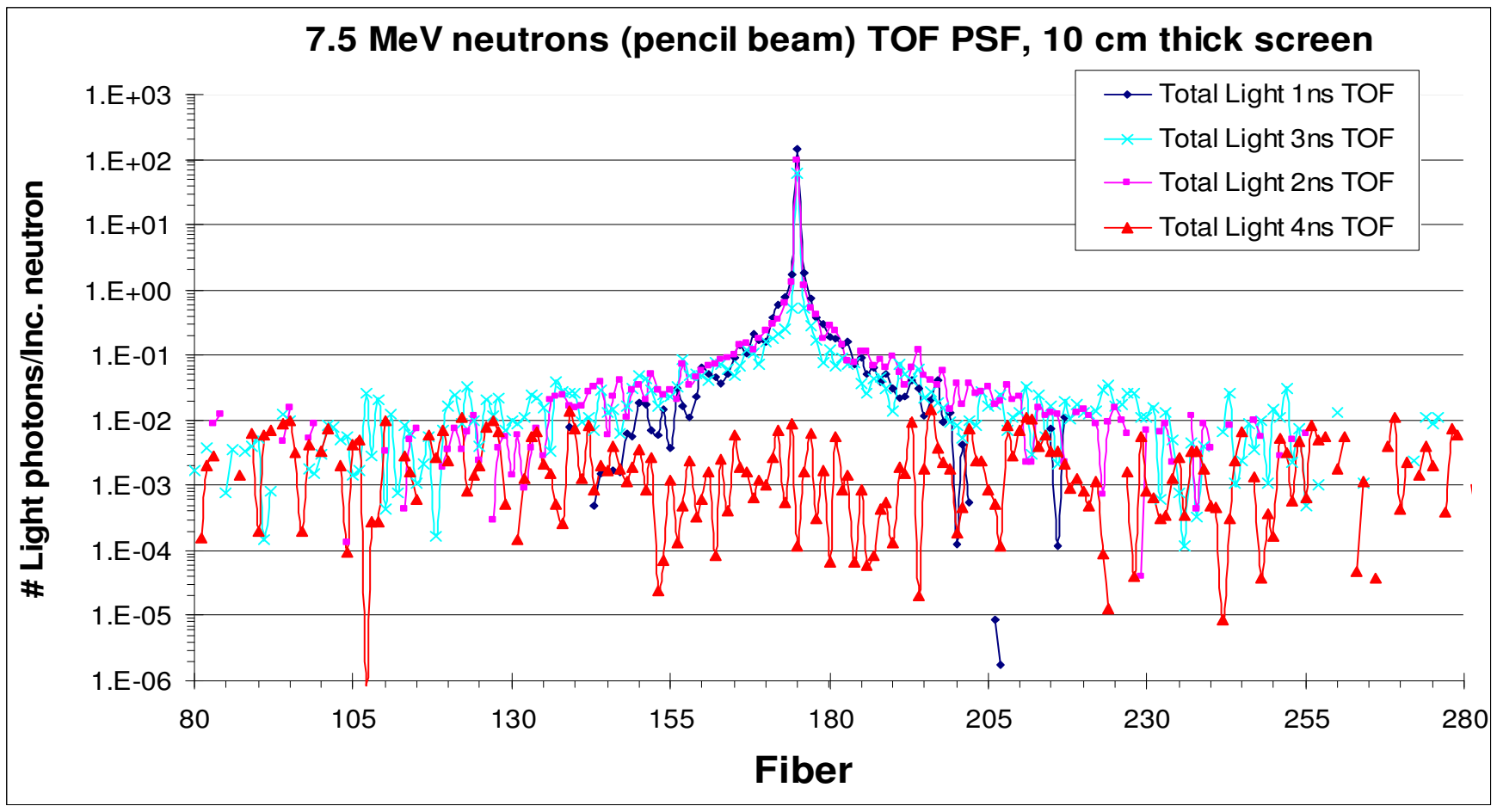

**Fig. 10 PSF vs. time for 7.5 MeV neutrons traversing 10 cm thick scintillator screen**

## 2.4 Reconstruction of temporal resolution

As shown in Fig. 2, TRION suffers from considerable loss of temporal contrast. We have tried to remedy this problem by deconvolving the temporal broadening function from the measured time spectrum. In addition to the instrumental response function one has now also to consider the intrinsic width and time structure of the radiation burst at the neutron producing target.

As a kernel or temporal point spread function (t-PSF), we used the gamma ray peak (produced by 12 MeV deuteron pulses on a thick $^{9}$Be reaction) measured with TRION. The gamma ray pulse represents already the time structure of the initial neutron burst, as all gamma-rays arrive at the detector with the same velocity and therefore the same TOF. In plastic scintillators, the scintillator response to the gamma-rays is a good approximation for the response to neutrons since for these, the scintillation decay time-constant for gamma-rays (internally-produced electrons) is not significantly different from that of neutrons (internally-produced protons). Hence, the gamma-peak measured by the detector contains most of the effects which broaden also the neutron time spectrum, namely, initial temporal width and structure of the radiation burst, the scintillator decay time and I-I gate width. Not included are of course the neutron scattering effects, which are relatively small compared to the previous ones mentioned. Fig. 11 shows the gamma-ray TOF peak as measured by TRION (normalized to its area), used as an overall t-PSF of the system.

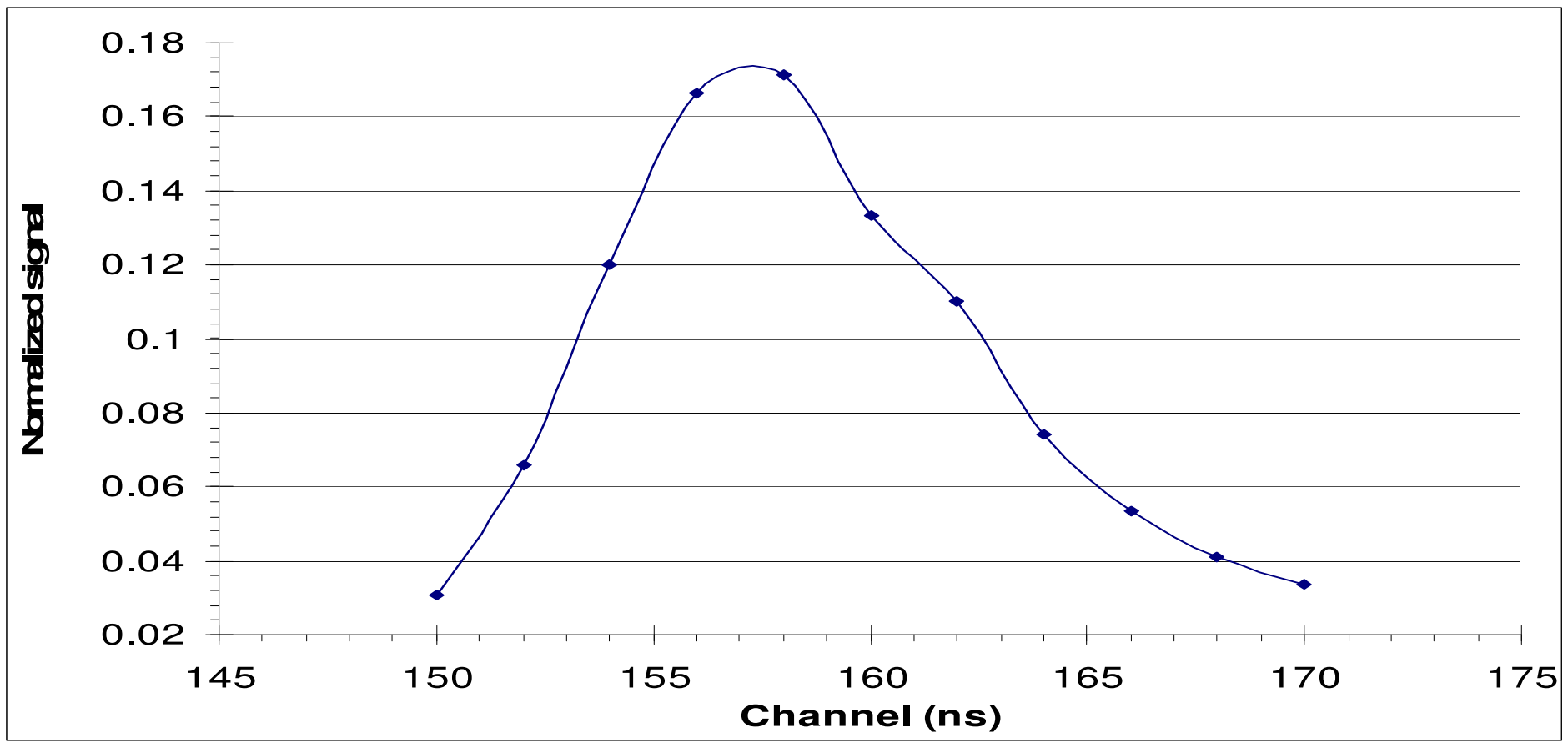

**Fig. 11 The gamma-ray TOF peak as measured by TRION**

Figure 12 shows the results of deconvolution (using the Lucy-Richardson algorithm) employing the gamma-ray TOF peak of Fig. 11 as a temporal PSF.

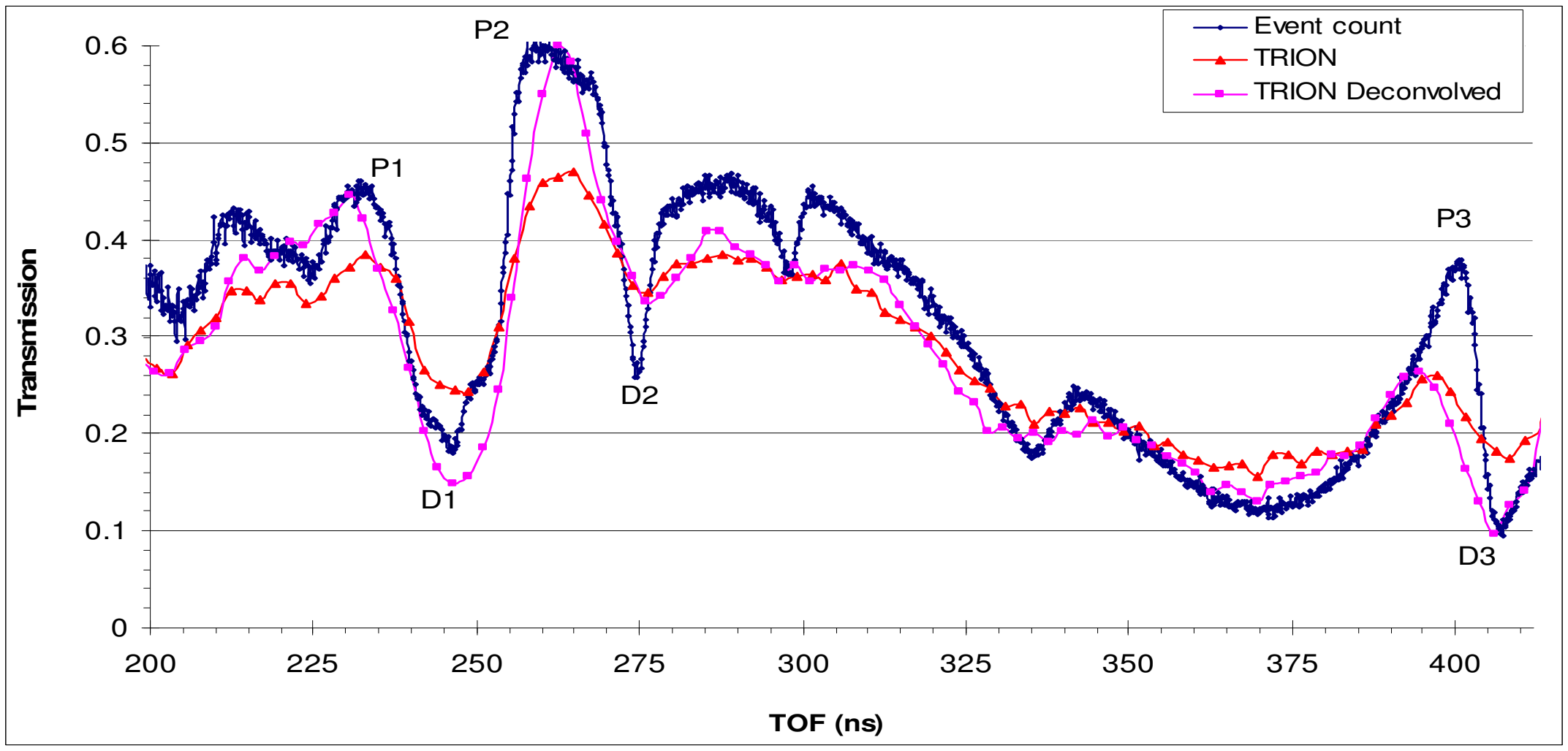

**Fig. 12 Transmission through 10 cm graphite, as measured by TRION Gen.2 before and after temporal deconvolution compared to calculated transmission**

Clearly, the temporal resolution has been recovered for most peaks of interest.

Table 1 shows the peak and dip values before and after deconvolution for the three peaks and dips indicated on Figure 10 by P or D, respectively.

**Table 1 Values of peaks and dips in transmission spectrum through graphite, before and after temporal deconvolution**

| | Expected Values | TRION - Before Deconvolution | TRION - Following Deconvolution | % Change Following Deconvolution |
|---|---|---|---|---|
| P1 | 0.403 | 0.384 | 0.445 | 15.90 |
| P2 | 0.527 | 0.469 | 0.599 | 27.70 |
| P3 | 0.372 | 0.259 | 0.263 | 1.54 |
| D1 | 0.143 | 0.242 | 0.148 | 38.84 |
| D2 | 0.108 | 0.345 | 0.336 | 2.60 |
| D3 | 0.072 | 0.174 | 0.097 | 44.25 |

## 3. Summary

In this paper theoretical and experimental evaluation of the parameters which affect temporal resolution of the TRION system were presented.

For intensifier gating windows shorter than 10 ns, the scintillator decay time causes more serious loss in temporal contrast than intensifier gating windows. This situation, however, reverses for intensifier gating windows equal to or longer than 20 ns.

For a 3 cm this scintillator screen, the time required for light to decay to its 1/100 value is within our gating time (4-6 ns). Therefore, for screen of similar thickness this effect is not of major importance, however, it should be considered and corrected for in thicker screens or shorter gating times.

Some of the lost temporal contrast may be recovered by use of the gamma-ray peak emitted from the d-Be reaction. The recorded gamma-ray TOF peak can be employed as a kernel or temporal point spread function (PSF) in a deconvolution process, since it essentially represents the system response to the original burst structure of the radiation pulse.